\documentstyle[prl,aps,twocolumn,epsfig,floats]{revtex}
\newfont{\fds}{cmtt10 scaled 1000}
\newcommand{\ds}{
\mbox{\hspace*{-0.1ex}.\hspace*{-0.3ex}{\fds *}\hspace*{0.1ex}}} 
 
\newcommand{\be}{\begin{equation}}
\newcommand{\ee}{\end{equation}}
\newcommand{\bea}{\begin{eqnarray}}
\newcommand{\eea}{\end{eqnarray}}
\newcommand{\tri}[3]{
\left[\!\!\begin{array}{c}#1\vspace*{-0.6ex}\\
#2\vspace*{-0.6ex}\\#3\end{array}\!\!\right]}
\def\>{\rangle}
\def\<{\langle}
\def\e{\epsilon}
\def\g{\gamma}
\def\dag{\dagger}
\def\ot{\otimes}
\def\noin{\noindent}
\def\non{\nonumber}

\newcommand{\eq}[1]{Eq.~(\ref{eq:#1})}
\newcommand{\ov}[1]{\overline{#1}}
\begin{document}

\preprint{}


\title{Simulation and reversal of $n$-qubit Hamiltonians using Hadamard
matrices}

\author{Debbie W. Leung}

\address{\vspace*{1.2ex}
	{IBM TJ Watson Research Center, P.O. Box 218,  
         Yorktown Heights, NY 10598 \\[1.2ex]}
}
	
\date{\today}
\maketitle


\begin{abstract}

The ability to simulate one Hamiltonian with another is an important primitive
in quantum information processing.
In this paper, a simulation method for arbitrary $\sigma_z \otimes \sigma_z$
interaction based on Hadamard matrices (quant-ph/9904100) is generalized for
any pairwise interaction.
We describe two applications of the generalized framework.  
First, we obtain a class of protocols for selecting an arbitrary interaction
term in an $n$-qubit Hamiltonian.  This class includes the scheme given in 
quant-ph/0106064v2.
Second, we obtain a class of protocols for inverting an arbitrary, possibly
unknown $n$-qubit Hamiltonian, generalizing the result in quant-ph/0106085v1.

\end{abstract}


\section{Introduction}

An important element in quantum information processing is the ability to
efficiently convert a set of primitives, determined by the physical system, 
to perform the desired task.
In many physical systems, the primitives are ``local manipulations'' such as
fast single qubit operations that can easily be controlled, and a {\em given}
nonlocal system Hamiltonian that cannot be changed.
In this case, the desired task may be approximated or {\em simulated} by
interspersing the given Hamiltonian evolution with local manipulations.  The
resources of simulation include the amount of local manipulations and the
total operation time of the given Hamiltonian.

Such simulation was extensively studied in the context of NMR quantum
computation~\cite{Linden98,Leung99,Jones99} in which the naturally
occurring Hamiltonian cannot be controlled.
Reference~\cite{Leung99} presents a method based on Hadamard matrices 
to convert the evolution due to any Hamiltonian of the form
\be
	H_1 = \sum_{i<j} g_{ij} \, \sigma_z^{(i)} \ot \sigma_z^{(j)}
		+ \sum_{i} \omega_i \, \sigma_z^{(i)} 
\label{eq:h1}
\ee
to that of a particular term $g_{lm} \sigma_z^{(l)} \ot \sigma_z^{(m)}$ where
$\sigma_z^{(i)}$ is a Pauli matrix acting on the $i$-th qubit.\footnote{
A similar method was reported independently in Ref.~\cite{Jones99}.}  
The same protocol applies universally for all coefficients $g_{ij}$ and
$\omega_i$.
Other arbitrary evolutions can in turns be obtained by reduction to the
universality construction~\cite{Div95a,Barenco95b,Deutsch95}.
The simulation of the single term is exact, and does not require frequent
local manipulations.  A related task to stop the interaction is also
addressed.
The method aims at minimizing the amount of local operations, 
measured by the number of interval of free evolution or the 
number of single qubit gates required. 

A more general problem was addressed recently in Ref.~\cite{Dodd01}.  
A major step is to convert the Hamiltonian 
\be
	H_2 = \sum_{ij} \sum_{\alpha \beta} g_{ij \alpha \beta} \, 
		\sigma_\alpha^{(i)} \ot \sigma_\beta^{(j)}
		+ \sum_{i} \vec{r}^{\,(i)} \cdot \vec{\sigma}^{(i)} 
\label{eq:h2}
\ee
to a single term $g_{lm \gamma \eta} \sigma_\gamma^{(l)} \ot
\sigma_\eta^{(m)}$, where $\sigma_{\alpha=x,y,z}$ denote 
the Pauli matrices.  
The single term is then used to simulate the {\em dynamics} due to any 
other Hamiltonian 
\be
	H_2' = \sum_{ij} \sum_{\alpha \beta} g'_{ij \alpha \beta} \, 
		\sigma_\alpha^{(i)} \ot \sigma_\beta^{(j)}
		+ \sum_{i} \vec{r}\,'^{(i)} \cdot \vec{\sigma}^{(i)} 
\label{eq:h2p}
\ee
For each required accuracy level, both local and nonlocal resources are
analyzed.

Related problems were also discussed in Refs.~\cite{Wocjan01,Janzing01},
focusing on more specific types of given Hamiltonians and the operation time
for the simulation.  Bounds are derived for the operation time to use 
\be
	H_3 = \sum_{ij} \sigma_z^{(i)} \ot \sigma_z^{(j)}
\label{eq:h3}	
\ee
to simulate an arbitrary Hamiltonian $H'_2$ given by \eq{h2p}.  
Methods and the required operating times for a Hamiltonian $H$ to 
simulate its inverse evolution $e^{i H t}$ are also studied.\footnote{
Throughout this paper, $\hbar = 1$, and the time evolution due to a
Hamiltonian $H$ is given by $e^{-iHt}$.  Note the $-$ sign in the exponent.
The inverse evolution is given by $e^{iHt}$.  This notation follows from the
Schr\"odinger equation.}
This is done for the Hamiltonians $H_3$ in \eq{h3} and $H_4$ given by 
\be
	H_4 = \sum_{\alpha} d_\alpha  
		\sum_{ij} \sigma_\alpha^{(i)} \ot \sigma_\alpha^{(j)} 
\label{eq:h4}	
\ee
which is essentially the most general Hamiltonian for pairwise interaction 
with permutation symmetry.\footnote{The two-qubit case was independently
considered in Ref.~\cite{Bennett01}}

The general principle in these simulation schemes is to transform some
coupling terms to the desired form and to cancel out the rest, by
interspersing the free evolution with single qubit operations.
In this paper, we generalize the framework developed in Ref.~\cite{Leung99} to
apply to the more general given Hamiltonian in \eq{h2}.
Using this framework, we find a class of schemes that select a term
$\sigma_\gamma^{(l)} \ot \sigma_\eta^{(m)}$ from a Hamiltonian given by
\eq{h2}.  This class of schemes includes the method presented in
Ref.~\cite{Dodd01}.
We also consider time reversal using the generalized framework, and present a
class of protocols to reverse an {\em arbitrary} Hamiltonian given by \eq{h2}. 
They require an operation time $c3nt$ of $H_2$ to simulate the reversed
evolution $e^{i H_2 t}$ where $1 \leq c \leq 2$ and $c \approx 1$ for large
$n$.
The schemes are designed to be independent of the given Hamiltonian, and are
applicable even when $H_2$ is {\em unknown}.  This generalizes the results in
Ref.~\cite{Janzing01}.

After the initial submission of this paper, related work were independently
reported in Ref.~\cite{Stollsteimer01}.
%
%
Methods to stop the evolution due to a Hamiltonian given by \eq{h2}, 
and to select {\em all diagonal} coupling terms between a designated 
pair of qubits are given.  
Simplifications for diagonal couplings and higher order terms were considered.
Our framework to stop the evolution is closely related to, but subtly
different from that in Ref.~\cite{Stollsteimer01}, and allows more flexibility
in selecting coupling terms, such as the selection of any {\em single}
interaction term.
We are also interested in a tighter bound on the required local resources,
and in the inversion of Hamiltonians. 

This paper is structured as follows. 
In Section~\ref{sec:reviewhad}, we review the framework and various resulting
schemes in Ref.~\cite{Leung99}, with a slight change from the original NMR
based notations.
The framework is generalized for any Hamiltonian given by \eq{h2} in
Section~\ref{sec:newhad}.  
The first application to select individual coupling terms from the given
Hamiltonian is discussed in Section~\ref{sec:newsel}.  
The second application to simulate time reversal is discussed in
Section~\ref{sec:timerev} as a simple application.
The technical details of the construction are described in  
Appendices \ref{sec:sylvester} and \ref{sec:ghm}.  

\section{Selective coupling using Hadamard matrices -- a review}
\label{sec:reviewhad}

\subsection{Statement of the problem}

We review the method developed in Ref.~\cite{Leung99}. 
Consider an $n$-qubit system, evolved according to the Hamiltonian 
\begin{equation}
	H_z = \sum_{i<j} g_{ij} \sigma_z^{(i)} \otimes
	\sigma_z^{(j)}
\label{eq:dipolar}
\,, 
\end{equation} 
where $g_{ij}$ are arbitrary coupling constants. 
The goal is to evolve the system according to only one term of the 
Hamiltonian: 
\begin{equation}
	\sigma_z^{(i)} \otimes \sigma_z^{(j)} 
\label{eq:zzij}
\,, 
\end{equation} 
using single qubit operations.  We call this task ``selective
coupling.''  This is closely related to the task of stopping 
the evolution or ``decoupling.''
We first develop a framework for decoupling.  Then we construct decoupling
and selective coupling schemes using Hadamard matrices.  
%

\subsection{Decoupling scheme for two qubits} 
\label{sec:example} 

We motivate the general construction using the simplest example of decoupling
two qubits.
Let $U_t = e^{-i H_z t}$, where the Hamiltonian is given by 
$H_z = g_{12} \, t \, \sigma_z^{(1)} \! \otimes \sigma_z^{(2)}$. 
We use the shorthand $X^{(i)}$ for $\sigma_x^{(i)}$.
We also use the important identity
\be
	U e^{M} U^\dagger = e^{UMU^\dag} 
\label{eq:cbasis}
\ee
where $M$ is any bounded square matrix and $U$ is any unitary matrix 
of the same dimension.  
As the Pauli matrices anticommute, 
\begin{eqnarray}
	X^{(2)} \, U_t \, X^{(2)}
	& = & X^{(2)} \; 
	      	e^{-i \, g_{12} \, t~\sigma_z^{(1)} \otimes \, \sigma_z^{(2)}} 
	      	X^{(2)}   
\non
\\
	& = & e^{-i \, g_{12} \, t~\sigma_z^{(1)} \otimes \, (X^{(2)} 
		\sigma_z^{(2)} X^{(2)})}
\non
\\
	& = & e^{-i \, g_{12} \, t~\sigma_z^{(1)} \otimes \, (-\sigma_z^{(2)})}
		~~ = ~~U_t^{-1}
\non
\end{eqnarray}
Thus adding the gate $X^{(2)}$ before and after the
free evolution {\em reverses} it, and  
\be
	X^{(2)} \, U_t \, X^{(2)} \, U_t = I \,.
\label{eq:refocus0}
\ee
This illustrates how single qubit operations can transform the action of 
one Hamiltonian to another. 

Equation (\ref{eq:refocus0}) can be written to highlight some 
essential features leading to decoupling: 
\begin{equation}
	e^{-i \, g_{12} \, t \, (+ \sigma_z^{(1)}) \otimes (- \sigma_z^{(2)})} 
\times 	e^{-i \, g_{12} \, t \, (+ \sigma_z^{(1)}) \otimes (+ \sigma_z^{(2)})} 
\label{eq:refocus}
\,. 
\end{equation}
Each factor corresponds to a ``time interval'' of evolution.  
\\ \noindent {\bf 1.}  In each interval, each $\sigma_z^{(i)}$ acquires a $-$
or $+$ sign, according to whether $X^{(i)}$ are applied or not before and
after the time interval.
\\ \noindent {\bf 2.}  The bilinear coupling is unchanged (negated) when the
signs of $\sigma_z^{(1)}$ and $\sigma_z^{(2)}$ agree (disagree).
\\ \noindent {\bf 3.} Since the matrix exponents commute, negating the
coupling for exactly half of the total time is necessary and sufficient to
cancel out the coupling.

The crucial point leading to decoupling is that, the signs of
the $\sigma_z$ matrices of the coupled qubits, controlled by the $X$
gates, disagree for half of the total time elapsed.

\subsection{Sign matrix and decoupling criteria} 

\begin{figure}[ht]
\begin{center}
\mbox{\psfig{file=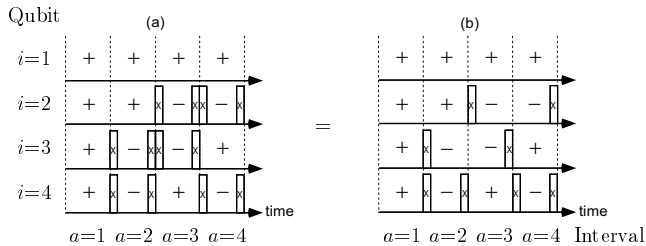,width=3.4in}}
\vspace*{1ex}
\caption{(a) Converting the sign matrix $S_4$ to the scheme in
Eq.~(\ref{eq:dec4}).  A ``$-$'' sign in the $i$-th row and $a$-th column
translates to $X^{(i)}$ (acting on the $i$-th qubit) before and after the
$a$-th time interval.
(b) Simplifying the scheme using $X^{(i)} X^{(i)} = I$.  
}
\label{fig:pulseseq}
\end{center}
\end{figure}

We now generalize the framework for decoupling to $n$ qubits.
Each of our schemes concatenates some $m$ equal-time intervals.
In each time interval, the sign of each $\sigma_z^{(i)}$ can be $+$ or $-$
as controlled by the $X$ gates.
Each scheme is specified by an $n \times m$ ``sign matrix'', with the $(i,a)$
entry being the {\em sign} of $\sigma_z^{(i)}$ in the $a$-th time interval.
The entries in each column represent the signs of all the qubits at the
corresponding time interval and the entries in each row represents the time
sequence of signs for the corresponding qubit.
We denote a sign matrix for $n$ qubits by $S_n$. 
For example, the scheme in Eq.~(\ref{eq:refocus}) can be represented by 
the sign matrix
\begin{equation}
	S_2 = \left[ \begin{array}{cc}
	{+}&{+}
\\	{+}&{-} 
	\end{array} \right] 
\label{eq:s2}
\,. 
\end{equation} 
Following the discussion in Sec.~\ref{sec:example}, we have 
\begin{quote} {\bf Decoupling~criteria -- version 1} ~~ 
Decoupling is achieved if any two rows in the sign matrix disagree in exactly
half of the entries.
\end{quote}
For example, the following sign matrix decouples four qubits: 
\begin{equation}
	S_4 = \left[ \begin{array}{cccc}
	{+}&{+}&{+}&{+}
\\	{+}&{+}&{-}&{-}
\\	{+}&{-}&{-}&{+}
\\	{+}&{-}&{+}&{-}
	\end{array} \right] 
\label{eq:s4}
\,. 
\end{equation} 
More explicit, the scheme is given by\footnote{ Note that the commuting
factors in Eq.~(\ref{eq:dec4}) are arranged to visually correspond to the sign
matrix.}
\be
	U_t  \times (X^{(3)} X^{(4)} \, U_t \, X^{(3)} X^{(4)})~ \times 
\hspace*{19ex}
\label{eq:dec4}
\ee
\vspace*{-6ex}
\bea
\nonumber
	~(X^{(2)} X^{(3)} \, U_t \, X^{(2)} X^{(3)})  
	\times (X^{(2)} X^{(4)} \, U_t \, X^{(2)} X^{(4)})  
\,, 
\eea
where $U_t = e^{-i H_z t}$ has six possible coupling terms in the $4$-qubit
Hamiltonian.  The relation between the scheme and $S_4$ is illustrated in
Figure \ref{fig:pulseseq}.

\noin {From} now on, we only consider the sign matrices, which 
completely represent the corresponding schemes.  
For $n$ qubits when $n$ is large, $n \times m$ sign matrices with small $m$
can be difficult to construct.  For example,
{\small
\begin{eqnarray}
\nonumber
	S_n = \left[ \begin{array}{cccccccccccc}
	{+}&{\cdots}&{+}&{+}&{\cdots}&{+}&{+}&{\cdots}&{+}&{+}&{\cdots}&{+}
\\	{+}&{\cdots}&{+}&{+}&{\cdots}&{+}&{-}&{\cdots}&{-}&{-}&{\cdots}&{-}
\\	{+}&{\cdots}&{+}&{-}&{\cdots}&{-}&{+}&{\cdots}&{+}&{-}&{\cdots}&{-}
\\	{ }&{\cdots}&{ }&{ }&{\cdots}&{ }&{ }&{\cdots}&{ }&{ }&{\cdots}&{ }
\\	{ }&{\cdots}&{ }&{ }&{\cdots}&{ }&{ }&{\cdots}&{ }&{ }&{\cdots}&{ } 
\\	{+}&{ -~ + }&{-}&{+}&{ -~ + }&{-}&{+}&{ - ~+ }&{-}&{+}&{ - ~+ }&{-}
	\end{array} \right] 
\,, 
\end{eqnarray} 
}

\vspace*{-1ex}
\noin in which an interval from a previous row is bifurcated takes 
$m = 2^{n-1}$. 
As the number of columns represents local resources for the simulation, our
goal is to find solutions with the smallest number of columns.  We solve this
problem by first rephrasing the decoupling criteria:

\begin{quote} {\bf Decoupling criteria -- version 2}~~ 
Identifying $\pm$ with $\pm 1$ in $S_n$, decoupling is achieved if any two
rows have zero inner product, or $S_n S_n^{T} = n I$.
\end{quote} 

\noin We now present very efficient solutions to the decoupling criteria,
which are derived from the Hadamard
matrices~\cite{CRC96,Sloanehp,vanLint92,MacWilliams77}.

\subsection{Hadamard matrices and decoupling scheme}
\label{sec:Hadamard}

A Hadamard matrix of order $n$, denoted by $H(n)$, is an $n \times n$ matrix
with entries $\pm 1$, such that
\begin{equation}
	H(n)H(n)^{T} = n I
\label{eq:ortho}
\,.
\end{equation}
Thus every $H(n)$, if exists, is a valid sign matrix corresponding to a
decoupling scheme for $n$ qubits using only $n$ time intervals.
The following is a list of interesting facts about Hadamard matrices (see
Refs.~\cite{CRC96,Sloanehp,vanLint92,MacWilliams77} for details and proofs).

\begin{enumerate} 
\item {\em Equivalence}~~Any permutation, or negation of any row or column of
a Hadamard matrix preserves the orthogonality condition.  Thus each Hadamard
matrix can be transformed to a {\em normalized} one, which has only $+$'s in
the first row and column.

\item {\em Necessary conditions}~~$H(n)$ exists only for $n = 1$, $n = 2$ or
$n \equiv 0 \bmod 4$.  

\item {\em Hadamard's conjecture}~\cite{Hadamard93}~~$H(n)$ exists for every
$n \equiv 0 \bmod 4$.  This famous conjecture is verified for all $n < 428$.  

\item {\em Sylvester's construction}~\cite{Sylvester67}~~If $H(n)$ and $H(m)$ 
exist, then $H(n) \otimes H(m)$ is a possible $H(nm)$.
In particular, $H(2^r)$ can be constructed as $H(2)^{\otimes r}$, which is
proportional to the matrix representation of the Hadamard transformation
for $r$ qubits.
\item {\em Paley's construction}~\cite{Paley33}~~Let $q$ be an odd prime
power.  If $q \equiv 3 \bmod 4$, then $H(q\!+\!1)$ exists; 
if $q \equiv 1 \bmod 4$, then $H(2(q\!+\!1))$ exists.  

\item {\em Numerical facts}~\cite{CRC96}~~
For an arbitrary integer $n$, let $\overline{n}$ be the smallest integer
satisfying $n \leq \overline{n}$ with {\em known} $H(\overline{n})$.
For $n\leq 1000$, $H(n)$ is known for every but $6$ possible orders, 
and $\overline{n} - n \leq 7$.  
For $n \leq 10000$, $H(n)$ is unknown for only $192$ possible orders and
$\overline{n} - n \leq 31$.
\end{enumerate}

\noin The nontrivial existence of so many Hadamard matrices may be better
appreciated by examining the following example of $H(12)$, obtained with
Paley's construction.
\begin{eqnarray}
\label{eq:h12}
	H(12) = \left[ \begin{array}{cccccccccccc}
	{+}&{+}&{+}&{+}&{+}&{+}& {+}&{+}&{+}&{+}&{+}&{+}
\\	{+}&{+}&{+}&{-}&{-}&{+}& {-}&{-}&{+}&{-}&{-}&{+}
\\	{+}&{+}&{+}&{+}&{-}&{-}& {-}&{+}&{-}&{+}&{-}&{-}
\\	{+}&{-}&{+}&{+}&{+}&{-}& {-}&{-}&{+}&{-}&{+}&{-}
\\	{+}&{-}&{-}&{+}&{+}&{+}& {-}&{-}&{-}&{+}&{-}&{+} 
\\	{+}&{+}&{-}&{-}&{+}&{+}& {-}&{+}&{-}&{-}&{+}&{-}
\\	{+}&{-}&{-}&{-}&{-}&{-}& {-}&{+}&{+}&{+}&{+}&{+}
\\	{+}&{-}&{+}&{-}&{-}&{+}& {+}&{-}&{-}&{+}&{+}&{-}
\\	{+}&{+}&{-}&{+}&{-}&{-}& {+}&{-}&{-}&{-}&{+}&{+}
\\	{+}&{-}&{+}&{-}&{+}&{-}& {+}&{+}&{-}&{-}&{-}&{+}
\\	{+}&{-}&{-}&{+}&{-}&{+}& {+}&{+}&{+}&{-}&{-}&{-} 
\\	{+}&{+}&{-}&{-}&{+}&{-}& {+}&{-}&{+}&{+}&{-}&{-}
	\end{array} \right] 
\,.
\end{eqnarray} 
Thus there is a simple decoupling scheme for $n$ qubits if $H(n)$ exists.
Using Hadamard matrices, decoupling and recoupling schemes for an arbitrary
number of qubits can be easily constructed, as will be shown next.

\subsection{Decoupling and selective coupling} 

\noin {\bf Decoupling} ~~ 
When an $H(n)$ exists, it corresponds to a decoupling scheme for $n$ qubits
concatenating only $n$ time intervals.
When an $H(n)$ does not necessarily exist, consider $H(\overline{n})$ and
choose any $n$ rows to form an $S_n$.  Then $S_n$ corresponds to a decoupling
scheme for $n$ qubits requiring $\overline{n}$ time intervals.
As an example, $S_9$ can be chosen to be the first nine 
rows of $H(12)$ in Eq.~(\ref{eq:h12}): 
\begin{eqnarray}
	S_9 = \left[ \begin{array}{cccccccccccc}
	{+}&{+}&{+}&{+}&{+}&{+}& {+}&{+}&{+}&{+}&{+}&{+}
\\	{+}&{+}&{+}&{-}&{-}&{+}& {-}&{-}&{+}&{-}&{-}&{+}
\\	{+}&{+}&{+}&{+}&{-}&{-}& {-}&{+}&{-}&{+}&{-}&{-}
\\	{+}&{-}&{+}&{+}&{+}&{-}& {-}&{-}&{+}&{-}&{+}&{-}
\\	{+}&{-}&{-}&{+}&{+}&{+}& {-}&{-}&{-}&{+}&{-}&{+} 
\\	{+}&{+}&{-}&{-}&{+}&{+}& {-}&{+}&{-}&{-}&{+}&{-}
\\	{+}&{-}&{-}&{-}&{-}&{-}& {-}&{+}&{+}&{+}&{+}&{+}
\\	{+}&{-}&{+}&{-}&{-}&{+}& {+}&{-}&{-}&{+}&{+}&{-}
\\	{+}&{+}&{-}&{+}&{-}&{-}& {+}&{-}&{-}&{-}&{+}&{+}
	\end{array} \right] 
\,.
\non
\end{eqnarray} 

\noin {\bf Selective coupling} ~~To implement selective coupling between the
$i$-th and the $j$-th qubits, any two rows in the sign matrix should be
orthogonal, except for the $i$-th and $j$-th rows that are identical.
The coupling $g_{ij} \; \sigma_z^{(i)} \otimes \sigma_{\rm z}^{(j)}$ acts
all the time while all other couplings are canceled.
The sign matrix can be obtained by taking $n-1$ rows from 
$H(\overline{n\!-\!1})$.
For example, to couple the last two among $9$ qubits, we can take $S_8$ to be
the $9 \times 8$ matrix obtained from appending the last row of $H(8)$ to
itself.
Alternatively, we can take the $2$-nd to the $9$-th rows of $H(12)$ in 
\eq{h12}, and repeat the last row: 
\begin{eqnarray}
\nonumber
	S_9 = \left[ \begin{array}{cccccccccccc}
	{+}&{+}&{+}&{-}&{-}&{+}& {-}&{-}&{+}&{-}&{-}&{+}
\\	{+}&{+}&{+}&{+}&{-}&{-}& {-}&{+}&{-}&{+}&{-}&{-}
\\	{+}&{-}&{+}&{+}&{+}&{-}& {-}&{-}&{+}&{-}&{+}&{-}
\\	{+}&{-}&{-}&{+}&{+}&{+}& {-}&{-}&{-}&{+}&{-}&{+} 
\\	{+}&{+}&{-}&{-}&{+}&{+}& {-}&{+}&{-}&{-}&{+}&{-}
\\	{+}&{-}&{-}&{-}&{-}&{-}& {-}&{+}&{+}&{+}&{+}&{+}
\\	{+}&{-}&{+}&{-}&{-}&{+}& {+}&{-}&{-}&{+}&{+}&{-}
\\	{+}&{+}&{-}&{+}&{-}&{-}& {+}&{-}&{-}&{-}&{+}&{+}
\\	{+}&{+}&{-}&{+}&{-}&{-}& {+}&{-}&{-}&{-}&{+}&{+}
	\end{array} \right] 
\,.
\end{eqnarray} 
The extra feature of this $S_9$ is that, all row sums are zero.  This is
because $H(12)$ in \eq{h12} is normalized, so that all rows except for the
first have zero row sums.  This automatically removes any local (linear) terms
$\sum_i \omega_i \sigma_z^{(i)}$ in the Hamiltonian, without extra local
resources (see Ref.~\cite{Leung99} for a full discussion).
Finally, we note that coupling terms involving disjoint pairs of 
qubits can be selected simultaneously.  

\subsection{Discussion}

\noin {\bf Upper bound on $\overline{n}$}~~
For $n$ qubits, selective coupling requires at most $\overline{n}$ intervals
and $n \overline{n}$ single qubit gates.
In fact, $\overline{n} = cn$ where $c$ is very close to the ideal lower bound 
$c=1$.  
First, if Hadamard's conjecture is proven, $\overline{n}$ only depends on 
$n \bmod 4$, and $\forall_n$ $\overline{n}\!-\!n \leq 3$.
Even without this conjecture, the present knowledge in Hadamard matrices
implies $\overline{n}\!-\!n \leq 8$ $\forall$ $n \leq 1000$, and 
$\overline{n} \!-\! n \leq 32$ $\forall$ $n \leq 10000$.
A detailed proof for $c \approx 1$ for {\em arbitrarily large} $n$ is given in
an Appendix of Ref.~\cite{Leung99}, while Sylvester's construction puts 
an immediate loose bound of $c < 2$. 
\\[1ex]
\noin {\bf Gate simulation vs dynamics simulation}~\cite{Bennett01} 
The previous discussion assumes that the goal is to simulate the final unitary
transformation due to the Hamiltonian $\sigma_z^{(i)} \! \otimes
\sigma_z^{(j)}$ for time $t$.
Due to the commutivity of all the possible coupling terms, we only need
to divide the time into $\overline{n}$ time intervals, each with {\em finite}
duration $t/\overline{n}$.
On the other hand, if the goal is to simulate the {\em dynamics} due to
$\sigma_z^{(i)} \! \otimes \sigma_z^{(j)}$ for time $t$, one should instead
apply the scheme to simulate the unitary gate ``$e^{-i \sigma_z^{(i)} \otimes
\sigma_z^{(j)} \Delta t}$'' $t/\Delta t$ times where $\Delta t$ is a small time
interval. 
We will focus on simulating the dynamics of a Hamiltonian in the rest of the 
paper.  

\section{Generalized framework for arbitrary $n$-qubit Hamiltonians} 
\label{sec:newhad}

We now generalize the framework for a more general given Hamiltonian 
(\eq{h2}): 
\be
	H_2 = \sum_{ij} \sum_{\alpha \beta} g_{ij \alpha \beta} \, 
		\sigma_\alpha^{(i)} \ot \sigma_\beta^{(j)}
		+ \sum_{i} \vec{r}^{\,(i)} \cdot \vec{\sigma}^{(i)} 
\label{eq:h22}
\ee
The goal is again to simulate the evolution due to one specific coupling term
$\sigma_\gamma^{(l)} \ot \sigma_\eta^{(m)}$.  
Passing from \eq{h1} to \eq{h22}, the first difference is the noncommutivity
of the terms in \eq{h22}.  The second difference is the presence of all three
Pauli matrices acting on the same qubit, besides a much larger number of
coupling terms.

We adopt a common approach~\cite{Dodd01,Wocjan01,Bennett01} that employs
sufficiently frequent local manipulations to make the effect of the
non-commutivity negligible.
This is based on the identity
\be	
	e^{-i K_1 t_1} e^{-i K_2 t_2} \approx e^{-i (K_1 t_1 + K_2
	t_2)} + {\cal O}(t_1 t_2) 
\,,
\label{eq:lie}
\ee
which implies that effects of non-commutivity are of higher order in the small
$t_i$ in \eq{lie}.  
Thus the discussion proceeds neglecting the non-commutivity, and products of
unitary evolutions are replaced by sums of the exponents.
With this simplification, the framework in the previous section is readily
generalized.

Again, we consider a class of schemes that concatenate (short) equal time
intervals of evolution.
The essential features for decoupling are as before: 
\\ \noindent {\bf 1.}  In each interval, each $\sigma_\alpha^{(i)}$ acquires a
$+$ or $-$ sign, which is controlled by the applied local unitaries to 
be described.  
\\ \noindent {\bf 2.}  The bilinear coupling $g_{ij \alpha \beta} \,
\sigma_\alpha^{(i)} \ot \sigma_\beta^{(j)}$ for $i \neq j$ is unchanged
(negated) when the signs of $\sigma_\alpha^{(i)}$ and $\sigma_\beta^{(j)}$
agree (disagree).
\\ \noindent {\bf 3.}  To the lowest order in the duration of the time
intervals, negating a coupling for exactly half of the intervals cancels it.
\\[1ex]
The generalized framework differs from the original one~\cite{Leung99} in
that, the signs of the three Pauli matrices $\sigma_{\alpha = x,y,z}^{(i)}$
acting on the same qubit $i$ in the same time interval are not 
independent.  In fact, the three signs multiply to $+$, because conjugating
$\vec r^{\,(i)} \cdot \vec \sigma^{(i)}$ by (local) unitaries on the $i$-qubit
transforms $\vec r^{\,(i)}$ by an SO(3) matrix.
Conversely, any sign assignment satisfying this constraint can be
realized.  The possible signs for $\sigma_x^{(i)}, \sigma_y^{(i)},
\sigma_z^{(i)}$ are $+++$, $+--$, $-+-$, $--+$, and are realized by 
applying $I^{(i)}$, $\sigma_x^{(i)}$, $\sigma_y^{(i)}$, $\sigma_z^{(i)}$
respectively before and after the interval.
Incorporating these considerations, we generalize the previous framework: 
\begin{quote}
A scheme for $n$ qubits that concatenates $m$ intervals can be specified by
three $n \times m$ sign matrices $S_x$, $S_y$, $S_z$, related by 
the {\em entry-wise} product $S_x\, \ds S_y = S_z$. 
The $(i,a)$ entry of $S_\alpha$ is the {\em sign} of $\sigma_\alpha^{(i)}$ in
the $a$-th time interval.  
\end{quote}

\noin We omit the number of qubits, $n$, in $S_x$, $S_y$, $S_z$ for simplicity
of notation.  The entry-wise product $\ds$ of two matrices is
also known as the Schur product or the Hadamard product.

\section{Selective coupling for $n$ qubits with arbitrary pairwise coupling}
\label{sec:newsel}

Under the generalized framework, we state the criteria for decoupling and 
selective coupling for $n$ qubits:
\begin{quote} {\bf Criteria for decoupling and selective coupling} ~~
Decoupling is achieved if any two rows taken from $S_x$, $S_y$, $S_z$ are
orthogonal.  Selective coupling of $\sigma_\gamma^{(l)} \ot \sigma_\eta^{(k)}$
is achieved if the $l$-th row of $S_{\gamma}$ is identical to the $k$-th
row of $S_{\eta}$, but any other pair of rows from $S_x$, $S_y$, $S_z$ are
orthogonal.  Local terms are removed if all row sums are zero.
\end{quote}

Sign matrices $S_{x,y,z}$ satisfying the criteria can be constructed from
special Hadamard matrices endowed with certain extra structures.
We now describe the constructions of the sign matrices, which elicit the
special properties required of the starting Hadamard matrices.  The more
difficult and technical constructions of these special Hadamard matrices are
given in detail in Appendices~\ref{sec:sylvester} and \ref{sec:ghm}.

Suppose we want to decouple $n$ qubits using a Hadamard matrix $H(m)$.
The orthogonality condition is automatically satisfied if the rows of
$S_{x,y,z}$ are taken to be distinct rows of $H(m)$.
It remains to ensure $S_x \ds S_y = S_z$.  
We call a set of $3$ vectors of equal length and with entries $\pm 1$ a
Schur-set if they entry-wise multiply to $++ \cdots +$.  For example, $\{
[+--], [-+-], [--+] \}$ is a Schur-set.
If $H(m)$ has at least $3n$ rows that {\em partition} into $n$ Schur-subsets,
one can ensure $S_x \ds S_y = S_z$ by choosing the $i$-th rows of $S_x, S_y,
S_z$ to be the rows of the $i$-th Schur-subset.
This poses the first extra property on $H(m)$ -- its rows partition into 
many Schur-subsets.
Note the immediate lower bound on the size of $H(m)$, 
$\lfloor (m-1)/3 \rfloor \geq n$ under this construction.  
This is not a strict lower bound, as a modification (to be described later) to
the construction enables $n$ to be replaced by $n+1$, but the modification 
removes some useful properties.
\vspace*{1ex}

Schemes for selective coupling can be derived from decoupling schemes as
follows.  Let $\{ \gamma, \eta, \nu \} = \{ x, y, z\}$ be distinct labels. 
To select the coupling $\sigma_\gamma^{(l)} \ot \sigma_\eta^{(k)}$
one can modify the decoupling sign matrices as follows:
\bea
\mbox{\psfig{file=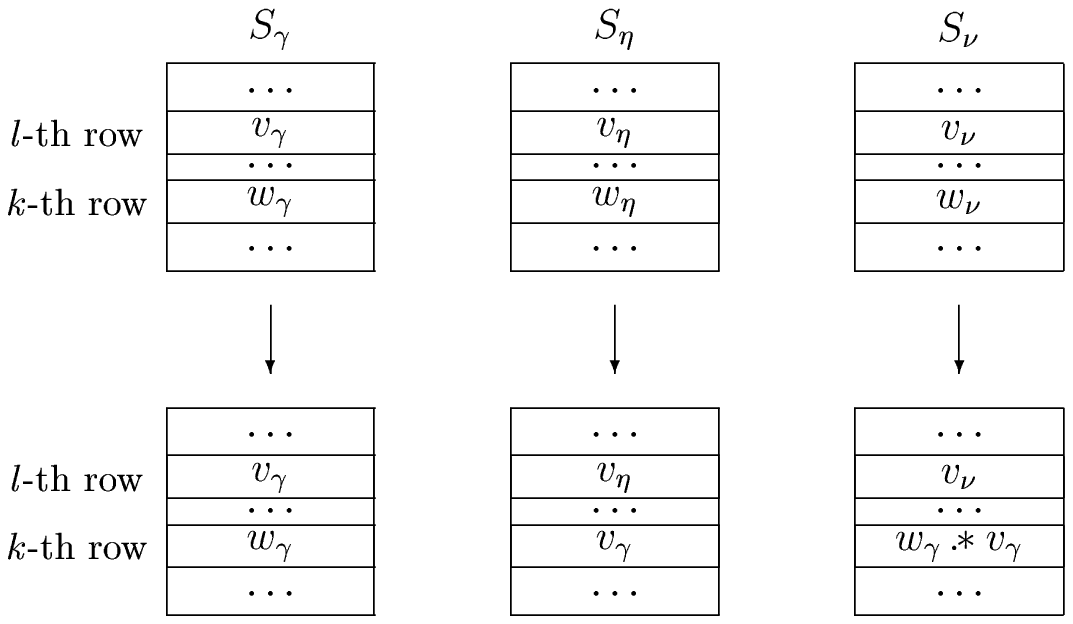,width=3.3in}}
\nonumber	
\eea
where $v_\alpha$, $w_\alpha$ for $\alpha \in \{ \gamma, \eta, \nu \}$ denote
the rows.
To select the coupling $\sigma_\gamma^{(l)} \ot \sigma_\gamma^{(k)}$, proceed
as before and further swap the $k$-th rows of $S_\gamma$ and $S_\eta$ 
(these are $w_\gamma$ and $v_\gamma$ in the lower diagram).  
In both cases, it is necessary to ensure $w_\gamma \ds v_\gamma$ is a row 
of $H(m)$ and is not used elsewhere in $S_{x,y,z}$. 
To ensure $w_\gamma \ds v_\gamma$ is a row of $H(m)$, one poses a
second special property in $H(m)$ that, it contains $5$ rows 
$f_{1, \cdots, 5}$ such that 
\be
	f_1 \ds f_2 = f_3 \ds f_4 = f_5 
\,.
\label{eq:extra2}
\ee  
Then we can simply choose $v_\gamma = f_5$, $v_{\eta,\nu} = f_{1,2}$ in the
decoupling scheme, and replace $w_\gamma$, $v_\gamma \ds
w_\gamma$ by $f_{3,4}$.
To ensure $f_{3,4}$ are not used elsewhere in $S_{x,y,z}$, the simplest method
is to exclude the Schur-subsets originally containing them.  
This is often unnecessary, as $f_{1,\cdots,5}$ can often be found such that 
$f_{3,4}$ are not in any Schur-subset in the starting decoupling scheme. 
Meanwhile, the $w_{x,y,z}$ can be ``recycled'' for an extra qubit.
Altogether, the scheme can handle $n \pm 1$ qubits depending on the situation.
%
%
%
%
%
\vspace*{1ex}

The above constructions of decoupling and selective coupling schemes 
involve only rows with zero sums if the Hadamard matrix $H(m)$ is normalized.  
This automatically removes all local (linear) terms without extra local
manipulations.   
It also means that, one can append an extra row $++\cdots+$ to all of 
$S_{x,y,z}$ to handle an extra qubit, the local terms of which, 
if exist, have to be dealt with outside of the scheme.  
\vspace*{1ex}

\begin{figure}[http]
\centerline{\mbox{\psfig{file=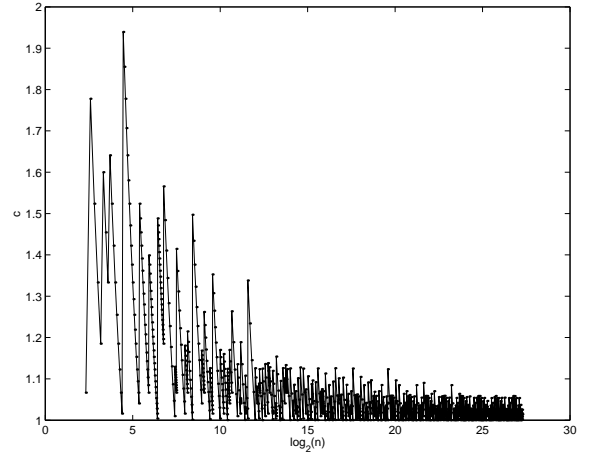,width=3in}}}
\caption{A plot of $c$ as a function of $n$, where $3cn$ is the number of
intervals for an $n$-qubit scheme.  The lower bound of $c$ is $1$.  
Using the Sylvester construction alone, $c$ fluctuates between $1$ and $2$
indefinitely.}
\label{fig:mplot}
\end{figure}

We now summary the results in Appendices \ref{sec:sylvester} and \ref{sec:ghm}
concerning the difficult issue of constructing normalized Hadamard matrices
that consist of many Schur-subsets and have $5$ rows satisfying \eq{extra2}.
The simplest of these are the Sylvester-type Hadamard matrices $H(2)^{\otimes
r}$ (referred to as Sylvester matrices from now on).
In Appendix~\ref{sec:sylvester}, we show how $H(2)^{\otimes r}$ can be
partitioned into $(2^r-1)/3$ and $(2^r-5)/3$ Schur-subsets when $r$ is even 
and odd respectively.  
When constructing a scheme for $n$ qubits based on Sylvester matrices, 
$r$ is chosen to satisfy the inequalities which are approximately\footnote{
The exact number depends on whether one is concerned with decoupling,
selective coupling, or inversion of Hamiltonians, and whether local terms are
to be handled by the scheme.  These may affect $n$ by a difference of 
$0$ or $\pm 1$. 
To avoid the cumbersome description of all possible variations, from now on,
we simply give approximates and omit $\pm 1$.  The readers can work out the
exact bounds tailored for their tasks. }
$2^{r-1} \leq 3n \leq 2^r$, and the number of intervals is approximately 
between $3n-6n$.
Appendix~\ref{sec:ghm} describes a more involved construction that combines a
{\em normalized generalized Hadamard matrix} with a Hadamard matrix having all
the required properties to produce a larger Hadamard matrix having the same
properties.
This significantly improves on the worst case number of intervals.   
For asymptotically large $n$, the number of intervals, $3cn$, has 
$c \approx 1$.  The value of $c$, as a function of $n$, is plotted in 
Figure \ref{fig:mplot}.
\vspace*{1ex}

In many applications, it may be useful to select more than one coupling term
from the Hamiltonian.  For example, one may select all coupling terms between
the $i$-th and the $j$-th qubits in $H_2$, by choosing the $i$-th and the
$j$-th rows of $S_{x,y,z}$ to be identical, and any other pairs of rows to be
orthogonal.  These $6$ identical rows have to be $++\cdots+$ in order to
satisfy $S_x \ds S_y = S_z$.  The required number of intervals is the same as
decoupling $n-2$ qubits.  This pairwise coupling can further be changed to a
desired form following methods in Refs.~\cite{Bennett01,Dodd01}.  Depending
on circumstances, this composite method can reduce the operation time of the
given Hamiltonian by a constant factor but can potentially increase the local
resources by a constant factor.  As a final remark, the selective coupling
schemes in both Refs.~\cite{Dodd01,Stollsteimer01} are to first select the
diagonal coupling terms or all coupling terms between a pair of qubits before
further manipulations.  The initial selection is done essentially using 
some version of the Sylvester construction, with even and arbitrary 
$r$ in Refs.~\cite{Dodd01} and \cite{Stollsteimer01} respectively.

\section{Universal Time reversal}
\label{sec:timerev}

We now apply the framework of sign matrices to simulate time reversal.  
As a first example, we apply the original framework in Ref.~\cite{Leung99} to
reverse $H_1$ given by \eq{h1}.  First construct a decoupling sign matrix
$S_n$ for $n$ qubits using a {\em normalized} Hadamard matrix
$H(\ov{n\!+\!1})$, excluding its first row.  In this case, all entries in the
first column of $S_n$ are ``$+$''.  A sign matrix for time reversal, denoted
by $T_{n}$, is obtained by removing the first column of $S_n$.
In $S_n$, any two rows have zero inner product, thus any two rows in $T_{n}$ 
have inner product $-1$, and any coupling term is reversed by exactly the same 
amount (note the significance of the $-$ sign in the inner product).
Furthermore, each row of $S_n$ has zero row sum, so, each row in $T_{n}$ has
row sum $-1$, and the local terms are reversed by the same amount as
well.
Therefore, any Hamiltonian given by \eq{h1} can be reversed.  
The simulation requires $n_I = \overline{n\!+\!1}-1$ intervals, and simulates
the reversal for the duration of one interval.  Hence the {\em simulation
factor}~\cite{Bennett01} is $1/n_I$ or the overhead~\cite{Wocjan01,Janzing01} 
is $n_I$.  The overhead is between $n$ and $n+3$ if
Hadamard's conjecture is true, and is $c n$ for $c \approx 1$ in any case.
Without local terms, $n_I = \overline{n}-1$ ranges between $n-1$ and $n+2$ if
Hadamard's conjecture is true.
Note that the protocol is independent of $H_1$, and can be applied to any 
$H_1$ which can even be unknown.
We remark that a lower bound $n-1$ was derived for the much more specific and
known Hamiltonian $H_3$ in \eq{h3}~\cite{Janzing01}.  In view of this, the
generality of the present result comes almost for ``free''.

To reverse a Hamiltonian given by \eq{h2} using the generalized framework, we
again construct sign matrices $S_{x,y,z}$ for a decoupling scheme (Section
\ref{sec:newsel}) with zero row sums.  Furthermore, our constructions can be
made to ensure all entries in the first columns of $S_{x,y,z}$ are $+$.
Therefore, removing the first columns in $S_{x,y,z}$ results in a time
reversal scheme.  The overhead is again the number of intervals in
the reversal scheme, which is $3cn$ for $c \approx 1$ for large $n$.
Again, the same protocol applies to any $H_2$ and thus applies to unknown 
Hamiltonians as well. 

As a comparison, the reversal method reported in Ref.~\cite{Janzing01} for
$H_4$ given by \eq{h4} depends on the knowledge of $d_\alpha$, and whether
they have the same sign or not.  When all $d_\alpha$ have the same sign, the
overhead is ${(n-1) |d_x + d_y + d_z| \over \max |d_\alpha|}$ 
or ${n |d_x + d_y + d_z| \over \max |d_\alpha|}$, which ranges from 
$\approx n-3n$.  
The generalized Hadamard matrix framework proposed can invert the much more
general Hamiltonian in \eq{h2} with simulation factor only slightly larger
than $3n$, without knowledge of the given Hamiltonian.  

\section{Conclusion} 

We have generalized the framework for Hamiltonian simulation and the methods
for decoupling and selective coupling in Ref.~\cite{Leung99}.
We rederive, as a special case of our construction, the crucial step of
selecting a coupling term in the simulation of $n$-qubit Hamiltonians in
Ref.~\cite{Dodd01}.
We also apply the technique to extend the time reversal problem considered in
Ref.~\cite{Janzing01} from permutation invariant purely nonlocal Hamiltonians
to an arbitrary $n$-qubit Hamiltonian. 

Our framework based on sign matrices allows the complicated criteria for
various simulation tasks to be rephrased in relatively simple orthogonality
conditions, for which solutions can be obtained with the connections to
Hadamard matrices.

\section{Acknowledgments}

This generalization was inspired by the work presented in Ref.~\cite{Dodd01}.
We are indebted to Dominik Janzing, Marcus Stollsteimer, and Pawel Wocjan for
pointing out a critical mistake in the initial version of the paper.  The
construction in Appendix \ref{sec:ghm} in the second version was partly
inspired by the mention of OA(48,13,4,2) in Ref.~\cite{Stollsteimer01}.
We thank David DiVincenzo, Aram Harrow, and Barbara Terhal for helpful
discussions and suggestions for the paper.
We thank Robin Huang, Jim Leonard, and Kathleen Falcigno for providing 
the author with timely access to some important references.
DWL is supported in part by the NSA and ARDA under the US Army Research
Office, grant DAAG55-98-C-0041.

 

\appendix

\section{Schur-subsets in Sylvester matrices}
\label{sec:sylvester}

In this Appendix, we study the properties of the Sylvester matrices useful 
for decoupling, selective coupling or inversion of Hamiltonian.  
First of all, they are always normalized.  
We now describe a method to partition the rows of the Sylvester
matrix $H(2)^{\otimes r}$ into $(2^r-1)/3$ and $(2^r-5)/3$ Schur-subsets
when $r$ is even and odd respectively.
It will be obvious from the construction that each $H(2)^{\otimes r}$ contains 
$5$ rows satisfying \eq{extra2}.  
\vspace*{1ex} 

First, we introduce some notations.  Let $\{0,1\}$ be the index set for the
rows and columns of $H(2) = \left[ \! \begin{array}{rr} 1 & \! \! \!1 \\ 1 &
\!\!-1 \end{array} \! \right]$.  For example, the $(0,0)$ entry is $1$.  We
use the shorthand $H_{ij}$ for the $(i,j)$ entry of $H(2)$.  Therefore $H_{ij}
= (-1)^{ij}$.
Likewise one can label the rows and columns of $H(2)^{\otimes r}$ with 
composite indices ${\bf i} = (i_1,i_2,\cdots,i_r)$ which are $r$-bit strings.
We have 
\bea
H_{\bf ij} &=& H_{i_1 j_1} \times H_{i_2 j_2} \times \cdots \times H_{i_r j_r} 
\nonumber
\\	   &=& (-1)^{i_1 j_1 + i_2 j_2 + \cdots + i_r j_r} 
\nonumber
\\	   &=& (-1)^{{\bf i} \cdot {\bf j}}
\label{eq:hcorr}
\eea
where ${\bf i} \cdot {\bf j}$ denotes the usual inner product of ${\bf i}$ 
and ${\bf j}$.   
For each ${\bf l}$,
\bea
 	 H_{\bf il} \times H_{\bf jl} \times H_{\bf kl} 
	= (-1)^{{\bf i} \cdot {\bf l} + {\bf j} \cdot {\bf l} 
		+ {\bf k} \cdot {\bf l}} 
	= (-1)^{({\bf i} + {\bf j} + {\bf k}) \cdot {\bf l}} \,.  
\eea
Therefore, the ${\bf i}$-th, ${\bf j}$-th, ${\bf k}$-th
rows form a Schur-set iff $\forall_{\bf l}$  
$H_{\bf il} \times H_{\bf jl} \times H_{\bf kl} = 1$ 
iff ${\bf i} \oplus {\bf j} \oplus {\bf k} = {\bf 0}$.
We refer to such a triple of $r$-bit strings as a Schur-set also. 
The problem reduces to showing that, the set of {\em non-zero} $r$-bit strings
partitions into $(2^{r}-1)/3$ and $(2^{r}-5)/3$ Schur-subsets when $r$ is even
and odd respectively.  
This can be proved by separate inductions on the even and odd values of $r$.
\vspace*{1ex}

For even values of $r \geq 2$, the induction hypothesis (IH) is that, the set
of all $r$-bit strings partitions into $K = (2^r - 1)/3$ Schur-subsets $E_1$,
$E_2$, $\cdots$, $E_K$, and the singleton $\{ {\bf 0} \}$.
The IH is clearly true when $r=2$.
Suppose it is true for some even $r \geq 2$.  
For each Schur-set $E_i = \{i_1,i_2,i_3\}$ of $r$-bit strings, we can obtain
$4$ Schur-sets of $(r\!+\!2)$-bit strings:
\bea
\nonumber
	\{01i_1, 10i_2, 11i_3\} 
\\
\nonumber
	\{01i_2, 10i_3, 11i_1\}
\\
\nonumber
	\{01i_3, 10i_1, 11i_2\}
\\
\label{eq:one2four}
	\{00i_1, 00i_2, 00i_3\}
\eea
We also have an addition Schur-set $\{ 01{\bf 0}, 10{\bf 0}, 11{\bf 0} \}$. 
Altogether, we find $4 (2^r - 1)/3 + 1 = (2^{r+2} - 1)/3$ Schur-sets of
$(r\!+\!2)$-bit strings that include all strings except for $00{\bf 0}$.
This completes the induction when $r$ is even.
For odd values of $r \geq 3$, the IH is: 
\begin{enumerate} 
\item the set of all $r$-bit strings partitions into $K = (2^r - 5)/3$
Schur-subsets $E_1$, $E_2$, $\cdots$, $E_K$, and a set of $5$ remainders $F =
\{w_1,w_2,w_3,w_4,{\bf 0}\}$, 
\item 
$E_K = \{k_1,k_2,k_3\}$ satisfies 
\bea 
	k_1 = w_1 \oplus w_2 = w_3 \oplus w_4 \,,
\nonumber
\\  
	k_2 = w_1 \oplus w_3 = w_2 \oplus w_4 \,. 
\label{eq:ih}
\eea
\end{enumerate}
For $r=3$, the IH is true, for example, by putting 
$E_{1 = K} = \{001, 100, 101\}$, and $F = \{010,011,110,111,{\bf 0}\}$. 
Suppose the IH is true for some odd $r \geq 3$.   
Using the method of \eq{one2four}, we obtain $4(K-1)$ Schur-subsets of
$(r\!+\!2)$-bit strings from $E_1$, $\cdots$, $E_{K-1}$.
There are $32$ remaining strings, $\{00,01,10,11\} \times
\{k_1,k_2,k_3, w_1,w_2,w_3,w_4, {\bf 0}\}$.  If we represent distinct 
bit strings as distinct points, and Schur-sets as triangles, the IH 
implies the following relations for the $6$ $r$-bit strings: 
\be
\mbox{\psfig{file=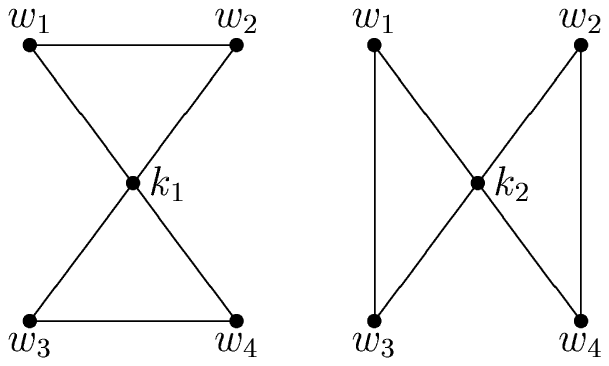,width=1.62in}}
\label{eq:h8}
\ee
Note that two triangles that share a common vertice represent 
two Schur-sets that are not disjoint.  
{From} \eq{h8}, we can easily find another $8$ disjoint Schur-subsets of 
$(r\!+\!2)$-bit strings: 
\be
\mbox{\psfig{file=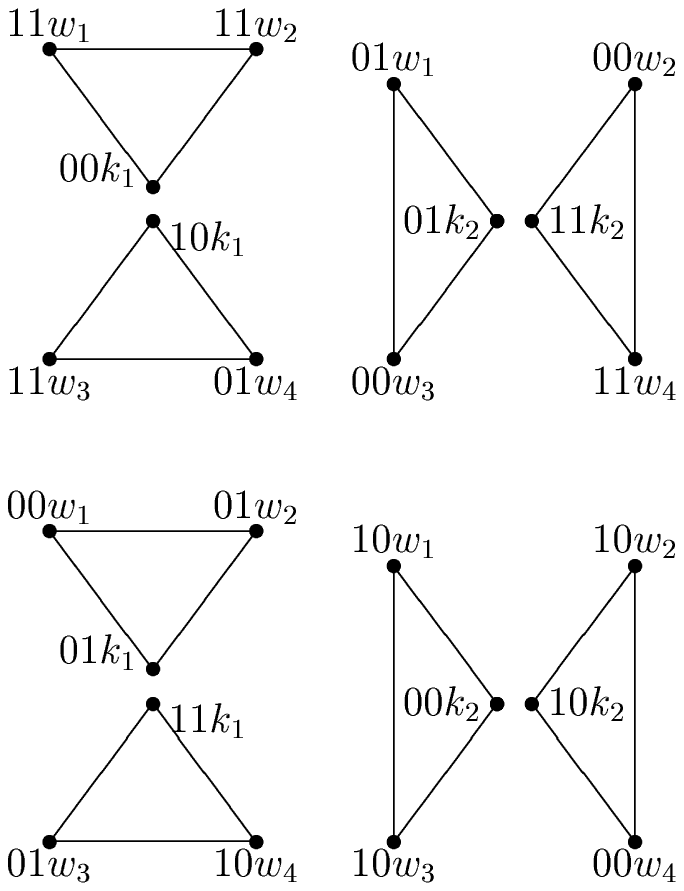,width=1.8in}}
\label{eq:h82}
\ee
Denote the remaining $8$ strings as $k'_1 = 11 {\bf 0}$, $k'_2 = 01 k_3$,
$k'_3 = 10 k_3$, $w'_1 = 00 k_3$, $w'_2 = 11 k_3$, $w'_3 = 01 {\bf 0}$, $w'_4
= 10 {\bf 0}$, ${\bf 0'} = 00 {\bf 0}$.
An additional Schur-set $\{k'_1, k'_2, k'_3\}$ can be formed, and the $5$
remaining rows $w'_{i=,1,2,3,4}, {\bf 0'}$ cannot form any more Schur-set.
Moreover, $k'_i, w'_i$ satisfy \eq{ih}.
Finally, the number of $(r\!+\!2)$-bit Schur-sets is 
$K' = 4 ((2^r - 5)/3 - 1) + 9 = (2^{r+2} - 5)/3$, completing the induction. 
\vspace*{1ex} 

The inductive proof, together with the identification $H_{{\bf ij}} =
(-1)^{{\bf i} \cdot {\bf j}}$, provides a constructive method to partition the
rows of $H(2)^{\otimes r}$ into Schur-subsets.
\vspace*{1ex} 

Finally, the fact $H^{\otimes r}$ contains $5$ rows satisfying \eq{extra2} is
manifest for odd $r$ in \eq{ih}.  In fact \eq{extra2} is exactly equivalent to
a pair of connected triangles as in \eq{h8}.  The case for even $r$ follows by
adding a ``$0$'' to each point in the joint triangles for odd $r$ in \eq{h8}.

\section{Construction using Generalized Hadamard matrices}
\label{sec:ghm}

In this Appendix, we construct Hadamard matrices of orders other than $2^r$
with the properties required of our schemes.  This is done by composing a
Hadamard matrix with a {\em generalized Hadamard matrix}~\cite{CRC96}.
\vspace*{1ex}

Let $G$ be a group of order $g$, with group operation $\star$. 
A generalized Hadamard matrix over $G$~\cite{CRC96,Hedayat99}, 
GH$(g,\lambda)$, is a $g \lambda \times g \lambda$ matrix whose 
entries $\g_{ij}$ are elements of $G$, and for 
$i \neq j$, the sequence $\{\g_{il} \star \g_{jl}^{-1}: 1 \leq l \leq g
\lambda\}$ contains each element of $G$ $\lambda$ times.
This sequence is the entry-wise ``division'' of the
$i$-th row by the $j$-th row.  
For example, a Hadamard matrix $H(4n)$ is a GH$(2,2n)$ over  
the multiplicative group $\{+1,-1\}$.
\vspace*{1ex} 

We are interested in generalized Hadamard matrices over GF(4), with $g=4$ 
elements written in an unusual manner: 
\be
\tri{+}{+}{+} \,, 
\tri{+}{-}{-} \,, 
\tri{-}{+}{-} \,, 
\tri{-}{-}{+} \,.  
\ee
The group operation $\ds$ is the entry-wise multiplication for the triples.  
The GH$(4,\lambda)$ is a $4 \lambda \times 4 \lambda$ array of triples 
(written as a column vector).
For example, a possible GH$(4,1)$ is given by 
{\small
\be 
	\g = \left[
	\begin{array}{cccc}
	\tri{+}{+}{+} & \tri{+}{+}{+} & \tri{+}{+}{+} & \tri{+}{+}{+} 
\vspace*{1ex}
\\
	\tri{+}{+}{+} & \tri{-}{-}{+} & \tri{+}{-}{-} & \tri{-}{+}{-} 
\vspace*{1ex}
\\
	\tri{+}{+}{+} & \tri{-}{+}{-} & \tri{-}{-}{+} & \tri{+}{-}{-} 
\vspace*{1ex}
\\
	\tri{+}{+}{+} & \tri{+}{-}{-} & \tri{-}{+}{-} & \tri{-}{-}{+} 
	\end{array}
	\right]
\label{eq:gamma}
\ee
}

\noindent We state some useful facts about generalized Hadamard
matrices (see for example, Ref.~\cite{CRC96} Part IV). 
\begin{enumerate}
\item {\em Equivalence}~\cite{Drake79}~~Permuting rows or
columns, or multiplying any row or column by a fixed element from the center
of $G$ preserve the defining properties of a GH$(g,\lambda)$.  Therefore, any
GH$(g,\lambda)$ over an abelian group $G$ is equivalent to a {\em normalized}
one, in which all entries in the first row and column are equal to the
identity element in $G$.

\item {\em Kronecker product}~\cite{Shrikhande64,Drake79}~~The Kronecker
product\footnote{
The Kronecker product in this case is defined with the group
operation $\star$ replacing the usual multiplication.} 
of GH$(g,\lambda_1)$ and GH$(g,\lambda_2)$ over the same abelian group $G$ is
a GH$(g, g \lambda_1 \lambda_2)$ over $G$.
\item \cite{deLauney86} Let $q=g \lambda - 1$ be a prime power.
If GH$(g,\lambda)$ over $G$ exists, $\forall_{t \geq 0}$   
GH$(g,\lambda q^t)$ over $G$ exists.   
\end{enumerate}
Applying the above facts to GF(4), the existence of GH$(4,\lambda)$ for
$\lambda = 1,2$, implies that for $\lambda = 1$, $2$, $3$, $4$, $8$, $9$,
$12$, $14$, $16$, $24$, $27$, $33$, $\cdots$.
\vspace*{1ex}

Let $H(m)$ be a Hadamard matrix with all the required properties: normalized,
having $5$ rows $f_i$ satisfying \eq{extra2}, and having some $3n$ rows
forming $n$ disjoint Schur-sets.  We can assume that the rows of $H(m)$ are
ordered so that members of each Schur-subset occur consecutively, starting
from the first row, and the last $m-3n$ rows do not belong to any
Schur-subsets.
For example, the $1$-st to $3$-rd rows form a Schur-subset, and same for 
the $4$-th to $6$-th rows, and so on.   
Converting every $3$ consecutive rows into a row of triples.
Do this for the first $3n$ rows of $H(m)$.  
The resulting array $\tilde{H}_E$ is an $n \times m$ array of triples.
Call the last $m-3n$ rows of $H(m)$ $H_F$. 
For example,
\bea
	H(2)^{\otimes 2} & = &    
	\left[ 
	\begin{array}{cccc}
	+ & - & + & - 
\\	+ & + & - & - 
\\	+ & - & - & +
\\	+ & + & + & + 
	\end{array}
	\right] 
\,,
\nonumber
\\
	\tilde{H}_E & = &
	\left[ 
	\begin{array}{cccc}
	\tri{+}{+}{+} & \tri{-}{+}{-} & \tri{+}{-}{-} & \tri{-}{-}{+} 
	\end{array}
	\right] 
\,,
\nonumber
\\
	H_F & = &
	\left[ 
	\begin{array}{cccc}
	+ & + & + & +
	\end{array}
	\right] 
\,.
\eea
As a second example, consider $H^{\otimes 4}$.  
The corresponding $H_F$ is $[+ \cdots +]$.  
Following the previous Appendix, Schur-subsets of $4$-bit strings are
$\{0101, 1010, 1111\}$ 
$\{0110, 1011, 1101\}$ 
$\{0111, 1001, 1110\}$ 
$\{0001, 0010, 0011\}$ 
$\{0100, 1000, 1100\}$.   
Thus, $\tilde{H}_E$ corresponding to $H(2)^{\otimes 4}$ is equal to 
{\footnotesize
\bea
\left[\rule{0pt}{28ex}
\begin{array}{c|c|c|c|c|c|c|c|c|c|c|c|c|c|c|c}
  + & - & + & - & + & - & + & - & + & - & + & - & + & - & + & - \\
  + & + & - & - & + & + & - & - & + & + & - & - & + & + & - & - \\
  + & - & - & + & + & - & - & + & + & - & - & + & + & - & - & + \\
    &   &   &   &   &   &   &   &   &   &   &   &   &   &   &   
\vspace*{-1.5ex} \\ \hline
    &   &   &   &   &   &   &   &   &   &   &   &   &   &   &   
\vspace*{-1.5ex} \\
  + & - & + & - & - & + & - & + & + & - & + & - & - & + & - & + \\
  + & + & - & - & + & + & - & - & - & - & + & + & - & - & + & + \\
  + & - & - & + & - & + & + & - & - & + & + & - & + & - & - & + \\
    &   &   &   &   &   &   &   &   &   &   &   &   &   &   &   
\vspace*{-1.5ex} \\ \hline
    &   &   &   &   &   &   &   &   &   &   &   &   &   &   &   
\vspace*{-1.5ex} \\
  + & + & - & - & - & - & + & + & + & + & - & - & - & - & + & + \\
  + & - & - & + & + & - & - & + & - & + & + & - & - & + & + & - \\
  + & - & + & - & - & + & - & + & - & + & - & + & + & - & + & - \\
    &   &   &   &   &   &   &   &   &   &   &   &   &   &   &   
\vspace*{-1.5ex} \\ \hline
    &   &   &   &   &   &   &   &   &   &   &   &   &   &   &   
\vspace*{-1.5ex} \\
  + & - & - & + & - & + & + & - & + & - & - & + & - & + & + & - \\
  + & - & + & - & + & - & + & - & - & + & - & + & - & + & - & + \\
  + & + & - & - & - & - & + & + & - & - & + & + & + & + & - & - \\
    &   &   &   &   &   &   &   &   &   &   &   &   &   &   &   
\vspace*{-1.5ex} \\ \hline
    &   &   &   &   &   &   &   &   &   &   &   &   &   &   &   
\vspace*{-1.5ex} \\
  + & + & + & + & - & - & - & - & + & + & + & + & - & - & - & - \\
  + & + & + & + & + & + & + & + & - & - & - & - & - & - & - & - \\
  + & + & + & + & - & - & - & - & - & - & - & - & + & + & + & + 
\end{array}
\right]
\nonumber
\eea
}

Returning to the general construction, we now compose a normalized $\g =
\mbox{GH}(g,\lambda)$ and $H(m)$ to form a new Hadamard matrix: 
\begin{enumerate}
\item Take the Kronecker product $\tilde{H}_E \otimes \gamma$ (under $\ds$).
\item Convert each row of triples in $\tilde{H}_E \otimes \gamma$ back into
$3$ rows of $\pm$ (an operation we call ``leveling'').  Call the resulting
matrix $H'_E$.  It is $3n 4 \lambda \times m 4 \lambda$.
\item Take the first coordinates of each triple in $\g$ and form a 
$4 \lambda \times 4 \lambda$ matrix of entries $\pm$.  
Take the (usual) Kronecker product of $H_F$ with the above matrix to obtain
$H'_F$.  It is $(m-3n) 4 \lambda \times m 4 \lambda$.
\item Append $H'_F$ to $H'_E$ to obtain an $m 4 \lambda \times m 4 \lambda$ 
matrix of $\pm 1$.  Call this $H'$.  
\end{enumerate}

We first show that the resulting matrix is indeed a Hadamard matrix.  Denote
by $\e_{ij}$ the $(i,j)$ entry of the {\em leveled} $\g$, and $h_{ij}$ the
$(i,j)$ entry of $H(m)$.  Then, $H'_E$ is explicitly given by:
\bea
\left[
\begin{array}{c|c|c}
  \tri{h_{11}}{h_{21}}{h_{31}} \ds \left[\rule{0pt}{3ex} ~~\g~~ \right] 
& \tri{h_{12}}{h_{22}}{h_{32}} \ds \left[\rule{0pt}{3ex} ~~\g~~ \right] 
& ~~ \cdots ~~ \\
& &  \vspace*{-1.5ex} \\ \hline & & \vspace*{-1.5ex} \\
  \tri{h_{41}}{h_{51}}{h_{61}} \ds \left[\rule{0pt}{3ex} ~~\g~~ \right] 
& \tri{h_{42}}{h_{52}}{h_{62}} \ds \left[\rule{0pt}{3ex} ~~\g~~ \right] &  \\
& &  \vspace*{-1.5ex} \\ \hline & & \vspace*{-1.5ex} \\
\dots & \cdots & ~~ \cdots ~~ \rule{0pt}{3ex}
\end{array}
\right]
\eea
or  
{\footnotesize
\bea
\left[
\begin{array}{ccc|ccc|c}
	h_{11} \, \e_{11} & h_{11} \, \e_{12} & \cdots 
	 & h_{12} \, \e_{11} & h_{12} \, \e_{12} & \cdots~ & ~~~\cdots~~~ \\
	h_{21} \, \e_{21} & h_{21} \, \e_{22} & \cdots~
	 & h_{22} \, \e_{21} & h_{22} \, \e_{22} & \cdots~ & ~~~\cdots~~~ \\
	h_{31} \, \e_{31} & h_{31} \, \e_{32} & \cdots~
	 & h_{32} \, \e_{31} & h_{32} \, \e_{32} & \cdots~ & ~~~\cdots~~~ \\
	& & & & & & \vspace*{-1.5ex} \\
	h_{11} \, \e_{41} & h_{11} \, \e_{42} & \cdots~
	 & h_{12} \, \e_{41} & h_{12} \, \e_{42} & \cdots~ & ~~~\cdots~~~ \\
	h_{21} \, \e_{51} & h_{21} \, \e_{52} & \cdots~
	 & h_{22} \, \e_{51} & h_{22} \, \e_{52} & \cdots~ & ~~~\cdots~~~ \\
	h_{31} \, \e_{61} & h_{31} \, \e_{62} & \cdots~
	 & h_{32} \, \e_{61} & h_{32} \, \e_{62} & \cdots~ & ~~~\cdots~~~ \\
	\vdots & \vdots & \ddots & \vdots & \vdots & \ddots & \\ 
	& & & & & & \vspace*{-1.5ex} \\ \hline
	& & & & & & \vspace*{-1.5ex} \\
	h_{41} \, \e_{11} & h_{41} \, \e_{12} & \cdots~ 
	 & h_{42} \, \e_{11} & h_{42} \, \e_{12} & \cdots~ & ~~~\cdots~~~ \\
	h_{51} \, \e_{21} & h_{51} \, \e_{22} & \cdots~
	 & h_{52} \, \e_{21} & h_{52} \, \e_{22} & \cdots~ & ~~~\cdots~~~ \\
	h_{61} \, \e_{31} & h_{61} \, \e_{32} & \cdots~
	 & h_{62} \, \e_{31} & h_{62} \, \e_{32} & \cdots~ & ~~~\cdots~~~ \\
	& & & & & & \vspace*{-1.5ex} \\
	h_{41} \, \e_{41} & h_{41} \, \e_{42} & \cdots~
	 & h_{42} \, \e_{41} & h_{42} \, \e_{42} & \cdots~ & ~~~\cdots~~~ \\
	h_{51} \, \e_{51} & h_{51} \, \e_{52} & \cdots~
	 & h_{52} \, \e_{51} & h_{52} \, \e_{52} & \cdots~ & ~~~\cdots~~~ \\
	h_{61} \, \e_{61} & h_{61} \, \e_{62} & \cdots~
	 & h_{62} \, \e_{61} & h_{62} \, \e_{62} & \cdots~ & ~~~\cdots~~~ \\
	\vdots & \vdots & \ddots & \vdots & \vdots & \ddots & \\ 
	& & & & & & \vspace*{-1.5ex} \\ \hline
	& & & & & & \\
	\cdots & \cdots & \cdots & \cdots & \cdots & \cdots & \\ 
	& & & & & & 
\end{array}
\right]
\label{eq:m}
\eea
}

\noindent {From} \eq{m}, the rows of $H'_E$ are of the form:
\be
	[h_{i1}, h_{i2}, \cdots~] \otimes [\e_{j1}, \e_{j2}, \cdots~ ] 
\ee
where $i \in \{1,\cdots,3n\}$, $j \in \{1,\cdots,12 \lambda \}$ and 
$i=j \bmod 3$.  
The rows of $H'_F$ are of the same form, with 
$i \in \{3n\!+\!1,\cdots,m\}$, $j = 3 k + 1$ for 
$k \in \{0,\cdots, 4 \lambda - 1\}$.  
Thus each row in $H'$ is specified by $i$ and $j$.   
Note that if $i \neq i'$, $[h_{i1}, h_{i2}, \cdots~]$ and $[h_{i'1}, h_{i'2},
\cdots~]$ are orthogonal.
If $j \neq j'$ and $j = j' \bmod 3$, then 
$[\e_{j1}, \e_{j2}, \cdots~ ]$ and $[\e_{j'1}, \e_{j'2}, \cdots~ ]$ 
are orthogonal since $\g$ is a generalized Hadamard matrix.
Therefore, any two distinct rows in $H'_F$ are orthogonal (orthogonal in at
least one tensor component), and any row from $H'_E$ is orthogonal to any row
from $H'_F$ (orthogonal in the first tensor component).  
Finally, let $i,j$ and $i',j'$ specify two distinct rows in $H'_E$.
If $i \neq i'$, they are orthogonal in the first tensor component. 
If $i=i'$, then $j \neq j'$, $j = j' \bmod 3$, and they are orthogonal in 
the second tensor component.  
Thus $H'$ has orthogonal rows and is a Hadamard matrix.\footnote{
Readers who are familiar with combinatorial designs may interpret the core
part of the above construction as taking the Kronecker product of an {\em
orthogonal array} ($\tilde{H}_E$) and a {\em difference matrix} ($\g$), both
over GF(4), to obtain a larger orthogonal array~\cite{CRC96,Hedayat99}.  The 
above construction is more elementary and allows slightly more features. }
\vspace*{1ex}  

Note that the rows of $H'_E$ completely partition into Schur-subsets due to
the group structure of the triples under $\ds$.
The ratio of the maximum number of qubits handled by the scheme to
the number of intervals is the same in $H'$ and $H(m)$ -- everything is 
rescaled by a factor of $4 \lambda$.
This construction is therefore most useful when $H(m)$ has a
large fraction of rows forming Schur-subsets, and when $\lambda$ is not 
a multiple of $4$.    
\vspace*{1ex}

One can verify that the above construction results in a normalized $H'$ 
up to a permutation of the rows.  Furthermore, because $\g$ is normalized 
$[h_{i1}, h_{i2}, \cdots~] \otimes [+ + \cdots~ ]$ 
occurs in $H'$ for all $i$, 
Therefore, $H'$ contains $5$ rows $f_i \otimes [+ + \cdots~ ]$ that satisfy 
\eq{extra2}.
\vspace*{1ex}  

Let the number of intervals required for an $n$-qubit scheme be $3cn$, and 
consider $c$ as a function of $n$.  
We now use the above construction to put a loose upper bound on $c$ 
for large $n$.  A tighter bound on $c$ for smaller values of $n$ is   
plotted in Figure~\ref{fig:mplot}.
A scheme formed by composing $H(m) = H(2)^{\otimes (r-2)}$ and
$\g = \mbox{GH}(4,3^{(t+1)})$ has $2^r 3^{(t+1)}$ intervals, and handles at
least $(2^r-20) 3^{\,t}$ qubits.
For a given $n$, we find the smallest value of $(2^r-20) 3^{\,t}$ larger than
$n$.  Let $n = 2^{r_o + \Delta}$ where $\Delta \in [0,1)$, and $\e =
\log_2(2^r/(2^r-20))$.  We look for the smallest non-negative value of
\be
	\log_2 ( (2^r-20) 3^t/n ) = r - \e + t \log_2 3 - r_o - \Delta
\label{eq:match} 
\ee
where $t$, $r$ are positive integers.
When $n$ is large, so is $r_o$.
Since $\log_2 3$ is irrational, $\{(t \log_2 3) \bmod 1 \}$ is a dense subset 
of $[0,1)$.  
We can find some $t \leq r_o/2\log_2 3$ such that $t \log_2 3 = N + \delta$,
$\delta \in [0,1)$, and $\delta - (\Delta + \e \bmod 1)$ is small and
nonnegative.
We also choose $r = r_o - N + \lfloor \Delta + \e \rfloor$.  Note that $r$ is
at least $r_o/2$.  Then,
\bea
	\log_2 c & = & \log_2 (2^{r} 3^{t+1}/3n) 
\nonumber
\\
	& = & r + t \log_2 3 - r_o - \Delta 
\nonumber
\\
	& = & r_o - N + \lfloor \Delta + \e \rfloor + N + \delta - r_o - \Delta
\nonumber
\\
	& = & \lfloor \Delta + \e \rfloor + \delta - \Delta - \e + \e
\nonumber
\\
	& = & [\delta - (\Delta + \e \bmod 1)] + \e
\eea
Both terms in the last line can be made small when $r_o$ is large, and 
$c \approx 1$.  

\end{document}